\documentclass{article}
\usepackage[utf8]{inputenc}

\usepackage{graphicx}
\usepackage{epstopdf}
\usepackage{subfigure}
\usepackage[utf8]{inputenc}
\usepackage{natbib}
\usepackage{amsmath}
\usepackage{amsfonts}
\usepackage{amssymb}
\usepackage{mathtools}
\usepackage{multirow}
\usepackage{appendix}
\usepackage{indentfirst}
\usepackage{hyperref}
\usepackage{color}
\usepackage{xcolor}

\usepackage{natbib}

\hypersetup{colorlinks=true,
	linkcolor=blue,
	filecolor=blue,      
	urlcolor=blue,
	citecolor=blue}

\newcommand{\bfy}{\mathbf{y}}
\newcommand{\bff}{\mathbf{f}}
\newcommand{\bfbeta}{\boldsymbol{\beta}}
\newcommand{\bfepsilon}{\boldsymbol{\epsilon}}
\newcommand{\bfkappa}{\boldsymbol{\kappa}}
\newcommand{\bfSigma}{\boldsymbol{\Sigma}}
\newcommand{\bfGamma}{\boldsymbol{\Gamma}}
\newcommand{\bfTheta}{\boldsymbol{\Theta}}
\newcommand{\bfDelta}{\boldsymbol{\Delta}}

\author{Vitor G. C. da Silva$^1$,
	Kelly C. M. Gon\c{c}alves$^1$, 
	and Jo\~ao B. M. Pereira$^1$}

\date{%
	$^1$Departamento de M\'etodos Estat\'{\i}sticos, Universidade Federal do Rio de Janeiro%
}

\title{Bayesian factor models for multivariate categorical data obtained from questionnaires}

\begin{document}

\maketitle

\begin{abstract}
Factor analysis is a flexible technique for
assessment of multivariate dependence and codependence. Besides being an exploratory tool used to reduce the dimensionality of multivariate data, it allows estimation of common factors that often have an interesting theoretical interpretation in real problems. However, standard
factor analysis is only applicable when the variables are scaled, which
is often inappropriate, for example, in data obtained from questionnaires in the field of psychology, 
where the variables are often categorical. In this framework, we propose a factor model for the analysis of multivariate ordered and non-ordered 
polychotomous data. The inference procedure is done under the Bayesian approach via Markov chain Monte Carlo methods. Two Monte-Carlo simulation 
studies are presented to investigate
the performance of this approach in terms of estimation bias, precision and assessment of the number of
factors. We also illustrate the proposed method to analyze participants' responses to the Motivational State 
Questionnaire dataset, developed to study emotions in laboratory and field settings.

{\it Keywords:} latent factors; data reduction; Metropolis-Hastings algorithm; categorical distribution; polychoric correlation

\end{abstract}

\section{Introduction}

Observations of categorical variables are very common in behavioral and social studies.
Typical examples are questionnaires in which individuals can be categorized according to how they classify themselves with respect to some issues. The answers to the questions are often supposed to be multiple indicators of one or more latent variables like, ability or attitude. On the other hand, if such multivariate polychotomous data are available, interest may be not only to investigate dependencies among them through latent factors, which allows determining how much the scale is related to the theoretical concepts that underlie it, but also to reduce dimensionality.

For example, the Perceived Stress Scale (PSS) is a 14-item scale that measures, using a five-point response differential, the degree to which the participants believe events in their lives are currently unpredictable, uncontrollable and overwhelming. Exploratory factor analysis in PSS analysis yields two-factors \citep{luft2007versao, hewitt1992perceived}, which besides allowing the production of reduced versions, provides two
important underlying factors, and their factor weights, with positive and negative interpretations, helping to construct a final score that explains an individuals's degree of stress. Another example is the Motivational State Questionnaire (MSQ), which has a four-point scale composed of 72 items, representing the full affective space \citep{revelle1998personality}. Exploratory factor analysis applied to MSQ yields a very clear two factor structure of energetic and tense arousal, although many of the words denote mixtures of these two constructs. Based on that, short alternative forms for the complete questionnaire were developed.

However, when the variables are continuous and measured on a common scale, principal component analysis and factor analysis often serve those purposes. In the analysis of binary or polychotomous variables, neither of these methods should be directly applicable, because they would neglect the nature of the data. Because the greatest stimulus for the development of models in this area has come from the field of educational testing, where the latent variables are conceived of as ``traits", it is usual to speak of latent trait models, when observed variables are discrete and latent variables are continuous \citep{bartholomew1999latent}. This contrasts with latent class models where the latent space is treated as categorical.

Some extensions of standard factor analysis are: \cite{wedel2003factor}, who developed a general class of factor-analytic models for the analysis 
of multivariate (truncated) count data, \cite{cagnone2012factor}, who proposed a latent variable model for binary data coming from an unobserved 
heterogeneous population and \cite{castro2015likelihood}, who considered censored non-normal random variables to deal with 
influential observations and censored data. Under the Bayesian paradigm, \cite{lopes2004bayesian} presented a Gibbs sampling algorithm to sample from the posterior distribution in standard factor analysis 
and estimate the number of latent factors, \cite{quinn2004bayesian} proposed a factor mixed model to deal with ordinal and continuous data, and 
\cite{zhang2014robust} developed a robust extension of the Gaussian factor
analysis model by replacing it for a multivariate t-distribution. In particular, for categorical variables, \cite{bartholomew1999latent} approached the problem from first 
principles and fundamentals without emphasizing computational techniques and not under the Bayesian paradigm.

Thus, the aim of this paper is to propose a factor analysis method with the Bayesian approach, for the case when all the variables are measured 
on an ordered or non-ordered categorical scale. For the ordered case, we suppose a latent process connected to the factor analysis structure, 
in which a continuous random vector is divided into intervals providing categories. In particular, we assume, 
conditionally on latent factor, the cumulative link model with 
probit and logistic link functions, which correspond, respectively, to assumption of normal and logistic distributions for the latent process 
errors. The intervals are defined by cutoff points, which can be assumed known or not. In this case, not only does the continuous vector provide 
a better interpretation for the model, it allows using well-known properties of the common normal factor model. In the non-ordered case, a 
categorical distribution (also called a generalized Bernoulli distribution) is assumed for the response vector and the factor structure is 
assumed for the logistic function of the response category probabilities. The inference is done under the Bayesian approach and
in particular we make use of Markov chain Monte Carlo (MCMC) methods to sample from the posterior distribution, since its kernel does not result 
in a known distribution.

The paper is organized as follows. Section \ref{sec2} presents the proposed models to handle, respectively, ordered and non-ordered categorical data. 
Some model properties are discussed together with the inference procedure, which follows the Bayesian paradigm and makes use of Metropolis-Hastings method, since the posterior full conditional distributions do not have an analytical closed form. Section \ref{sec3} presents the analysis of some 
synthetic datasets. First a Monte-Carlo simulation study is presented to evaluate the performance of the approach with respect to frequentist 
properties of the Bayes estimators. Then, two studies are presented with the purpose of compare model's fit under 
different numbers of latent factors. In Section \ref{sec4}, 
we illustrate the approach for the ordered category by analyzing the factor structure underlying emotions dataset collected by the Motivational State 
Questionnaire. Model comparison is performed using different criteria. Finally, Section \ref{sec5} discusses our main findings.

\section{Proposed models}\label{sec2}
\subsection{Model for ordinal data}\label{ordmodel}
Let ${\bf y}_i$ be an observed $p$-vector of polychotomous variables, such that $y_{ij}=k$ is an ordinal categorical variable indicating that the $j$-th response variable for sample unit $i$ is in category $k$. A possible way of modeling a categorized random variable $y_{ij}$ is to consider that it has been generated from a continuous latent variable, $y_{ij}^*$, divided into intervals whose bin boundaries are unknown. The categorical variable
is classified in category $k$ if, and only if, the associated continuous variable falls within, say, $\alpha_{j,k-1}$
and $\alpha_{j,k}$, that is
\begin{align*}
y_{ij} = k, \mbox{ if and only if }  \alpha_{j,k-1} < y_{ij}^{*} \leq \alpha_{j,k},  \; k = 1, \dots, K, 
\end{align*}
with $\alpha_{j,0} = -\infty$, or the lowest value that $y_{ij}^*$
can assume, and $\alpha_{j,K} = \infty$. Then one
can model the cumulative probability that the response variable lies in category $k$, or below it, at component $j$ and unit $i$ as:
\begin{align*}
\delta_{ijk}=P(y_{ij} \leq k) = P(y_{ij}^* \leq \alpha_{j,k}).
\end{align*}

Thus, we model the observed categories, treating the latent  process $y_{ij}^*.$  Moreover, we assume that the $p$-vector ${\bf y}_{i}^*$ is a
measurement of $q$ latent factors, ${\bf f}_i = (f_{1i}, \dots, f_{qi})'$ or, equivalently, the associations
among the observed variables are wholly explained by $q$ latent variables, with $q < p$, through the following factor model: for $i=1,\dots,n$
\begin{align}\label{factormodel}
\bfy_i^* = \bfbeta \bff_i +\bfepsilon_i, \mbox{ where}
\end{align}
(i) the factors ${\bf f}_i$ are independent with $\bff_i\sim N_q({\bf 0}_q,{\bf I}_q)$, for ${\bf I}_q$ an identity $q-$matrix and ${\bf 0}_q$ a zero $q-$vector, (ii) $\bfepsilon_i$ is an independent $p$-vector with scale parameter
$\bfSigma=diag(\sigma^2_1,\dots, \sigma_p^2)$, (iii) $\bfepsilon_i$ and ${\bf f}_s$ are independent for all $i$ and $s$, and (iv) $\bfbeta$ is the $p\times q$-factor loadings matrix.

In this work we assume two different models for $\bfepsilon_i$: 
\begin{align}
\bfepsilon_i &\sim N_p({\bf 0}_p,\bfSigma) \mbox { and } \label{normal_ord}\\
\bfepsilon_i &\sim Logistic_p({\bf 0}_p,\bfSigma). \label{logist_ord}
\end{align}

The cumulative probability that the response variable at component $j$ and unit $i$ lies in category $k$, or below it, is obtained conditionally  on the latent factor as:
\begin{align}\label{def_gamma}
\gamma_{ijk}= P(y_{ij}^* \leq \alpha_{j,k}\mid \bff_i).
\end{align}

Then, conditional on the latent factors, the normal model in (\ref{normal_ord}) implies that 
the implicitly assumed link function is a probit, so: for $i=1,\dots,n$, for $j=1,\dots,p$ and for $k=1,\dots,K$,
\begin{align}\label{gamma1}
\gamma_{ijk} = \Phi\left(\displaystyle\frac{\alpha_{j,k} - \sum_{l=1}^q \beta_{jl}f_{li}}{\sigma_j}\right),   
\end{align}
where $\Phi$ denotes the cumulative standard normal distribution.

On the other hand, conditional on the latent factors, the logistic model in (\ref{logist_ord}) implies implicitly 
a logit link function, this is: for $i=1,\dots,n$, for $j=1,\dots,p$ and for $k=1,\dots,K$,
\begin{align*}
\gamma_{ijk} = \frac{\exp\left(\displaystyle\frac{\alpha_{j,k} - \sum_{l=1}^q \beta_{jl}f_{li}}{\sigma_j}\right)}{1+\exp\left(\displaystyle\frac{\alpha_{j,k} - \sum_{l=1}^q \beta_{jl}f_{li}}{\sigma_j}\right)}.
\end{align*}

Since the inference procedure is done conditional on the latent factors and then integrating them out, it is not strictly necessary to obtain marginal properties 
of the models. Nevertheless, under the model (\ref{normal_ord}) and the normal assumption for the latent factors, we obtain the joint model
\begin{align}\label{jointnormal}
\left(\begin{array}{c}\bfy_i^*\\\bff_i\end{array}\right)\sim N_{p+q}\left[\left(\begin{array}{c}{\bf 0}_p\\{\bf 0}_q\end{array}\right),
\left(\begin{array}{cc}\bfbeta\bfbeta'+\bfSigma& \bfbeta\\\bfbeta'& {\bf I}_q\end{array}\right)\right],
\end{align}
for $\bfbeta$ and $\bfSigma$ defined in equation (1).
Under this specification, marginal on the latent factors $\bff_i$, the components of the $p$-vector ${\bf y_i}^\ast$ are not independent, which implies that the components of ${\bf y}_{i}$ are not independent either. The joint cumulative probability function of the response variable ${\bf y}_i$ is given by:
\begin{eqnarray*}
	\delta_{i{\kappa}} = P(y_{i1} \leq k_1,\dots,y_{ip} \leq k_p)= P(y_{i1}^\ast \leq \alpha_{1,k_1},\dots,y_{ip}^\ast \leq \alpha_{p,k_p}),
\end{eqnarray*}
where $\bfkappa = (k_1,\dots,k_p)$, with $k_l \in \{1,\dots,K\}$ and $l=1,\dots,p$. The marginal distribution of $y_{ij}$ is trivially obtained 
from (\ref{jointnormal}) and joint multivariate normal distribution
properties. In particular, 
the marginal link function in this case is also a probit, given by:
\begin{align*}
\delta_{ijk} = \Phi\left(\displaystyle\frac{\alpha_{j,k}}{\sqrt{ \sum_{l=1}^q \beta_{jl}^2+\sigma_j^2}}\right).   
\end{align*} 

However, assuming model (\ref{logist_ord}) makes analytical manipulation more difficult and numerical integration may be required 
in this case.

The normal assumption for the latent factors is a common choice in the standard factor analysis, mainly for 
facilitating analytical manipulations derived from multivariate normal distribution properties. In general, the latent factor distribution is changed in 
according to 
the error assumption. For example, to deal with a multivariate t-distribution, \cite{zhang2014robust} replaced the normality assumption
not only in the error distribution but also in the distribution of the latent factors, \cite{wedel2003factor} deal with Poisson distribution for the errors and 
showed that assuming normal and gamma distributions for the latent factors yield, under special cases, to known marginal distribution for the
response variable. In this work, although we only considered the normal distribution for the latent factors, taking advantage of the conjugation with
the model assumed in (\ref{normal_ord}) for the errors, other distributions could be assumed.

\subsubsection{Model properties}

In both models assumed for the error, the variance structure of the latent process distribution, given by:
\begin{align}\label{var_fact_mod}
\bfDelta=var({\bf y}_{i}^*\mid \bfbeta,\bfSigma)=\bfbeta\bfbeta'+\bfSigma,
\end{align}
is divided into a part explained by the common factors and the uniquenesses $\sigma^2_j$, $j=1,\dots,p$, which measure the residual variability in each of the data variables once that contributed by the factors is accounted for.

The models imply that, conditional on the common factors, the variables
are uncorrelated. Hence, the common factors explain all the dependence structure among the $p$ variables. Then, for any elements $y_{ij}^*$ and 
$y_{ij'}^*$ of $\bfy_i^*$, we have the characterizing moments: $var(y_{ij}^*\mid \bff_i) = \sigma^2_j$ , $cov(y_{ij}^*, y_{ij'}^* \mid \bff_i) = 0$, $var(y_{ij}^*) =\sum_{l=1}^q \beta_{jl}^2+\sigma_j^2$ and $cov(y_{ij}^*, y_{ij'}^*)=\sum_{l = 1}^{q} \beta_{jl} \beta_{j'l}.$ 

The proposed model suffers from two identifiability problems. One is due to the invariance of factor models under orthogonal transformations. 
This is, if one considers $\bfbeta^*= \bfbeta \bfGamma^{-1}$ and $\bff_i^*= \bfGamma \bff_i$, for any nonsingular matrix $\bfGamma$, 
the same model defined in (\ref{factormodel}) is obtained. The other is identifiability issues with the variance and bin boundaries associated 
with the latent variable distribution. Thus, the model must be further constrained to be free from identification problems. This is, the substitution of the parameters 
$\alpha_{j,k}^*=c\alpha_{j,k}$, $\sigma_j^*=c\sigma_j$, $\beta_{jl}^*=\sqrt{c}\beta_{jl}$ and 
$f_{li}^*=\sqrt{c}f_{li}$, for all $j=1,\dots,p$, $k=1,\dots,K$, $l=1,\dots,q$ and any $c>0$ implies that
\begin{align*}
\gamma_{ijk}^* = \Phi\left(\frac{c\alpha_{j,k}-c\sum_{l=1}^{q}\beta_{jl}f_{li}}{c\sigma_j}\right) = \gamma_{ijk},
\end{align*}
for $\gamma_{ijk}$ defined in equation (\ref{gamma1}).

To deal with the factor model invariance, the alternative adopted in this paper to identify the model is to constrain $\bfbeta$ to be a block lower 
triangular matrix, assumed to be of full rank, with strictly positive diagonal elements. This form provides both identification and often useful interpretation of the factor model \citep{lopes2004bayesian}. With respect to the identifiability issue related to the latent model, some authors suggest not performing inference about the variance of the latent random variable and cutoff points \citep{albert1993bayesian,de2000bayesian,higgs2010clipped}. Following this, since the variance is already constrained to deal with the factor model invariance, we just consider the cutoff points to be fixed, that is, $\alpha=\{\alpha_{j,k}, \mbox{ for all } j=1,\dots,p, k=1,\dots,K\}$.

\subsubsection{Inference procedure}\label{Inf1}

Let $\bfy = (\bfy_1,\dots,\bfy_n)$ denote a random sample with $n$ observations from the categorical variable and $\pi_{ijk}$
be  the  probability  that  the  response  variable  lies  in
category $k$ at component $j$ for unit $i$ conditional on ${\bf f}_i$, that is, $\pi_{ijk}=P(y_{ij}=k\mid {\bf f}_i).$ From equation (\ref{def_gamma}), it follows that $\pi_{ij1} = \gamma_{ij1},$ and $\pi_{ijk} = \gamma_{ijk}-\gamma_{ij,k-1},$ for $k=1,\dots,K.$

Let $\bfTheta= (\bfbeta,\bff_1,\dots,\bff_n,\sigma^2_1,\dots,\sigma^2_p)$
be the parameter vector. Following equation (\ref{factormodel}) and assuming for example the normal error in (\ref{normal_ord}), the likelihood function for $\bfTheta$ can be written as:
\begin{align*}
l(\bfTheta;{\bf y})&\propto \prod_{i=1}^n\prod_{j=1}^p\prod_{k=1}^K{\pi_{ijk}^{I(y_{ij} = k)}}\\
& = \prod_{i=1}^n\prod_{j=1}^p\prod_{k=1}^K{\left[\Phi\left(\displaystyle\frac{\alpha_{j,k} - \sum_{l=1}^q \beta_{jl}f_{li}}{\sigma_j}\right) - \Phi\left(\displaystyle\frac{\alpha_{j,k-1} - \sum_{l=1}^q \beta_{jl}f_{li}}{\sigma_j}\right)\right]^{I(y_{ij} = k)}} ,
\end{align*}
where $I(.)$ denotes the indicator function.

The inference procedure is performed under the Bayesian paradigm assuming the number of factors $q$ known, and model specification is complete after assigning a
prior distribution for $\bfTheta$, $p(\bfTheta)$. An advantage of following the Bayesian paradigm is that the inference procedure is performed under 
a single framework, and uncertainty about parameter estimation is naturally accounted for. We assume some
components of $\bfTheta$ are independent, {\it a priori}. More specifically,
\begin{align*}
p(\bfTheta)=\left[\prod_{i=1}^n{g(\bff_i)}\right]\left[\prod_{j=1}^p{p(\sigma^2_j)}\right]p(\bfbeta),    
\end{align*}
where $g(\bff_i)$ is the pdf of the $q$-multivariate normal with all the components of the mean vector equal to 0, and correlation identity matrix. 
We assume further that {\it a priori}, $\beta_{jl} \sim N(0,C_0)$, when 
$j\neq l$, $\beta_{jj} \sim N(0,C_0)I(\beta_{jj}>0)$ and $\sigma_j^{2}\sim IG(\nu/2,\nu s^2/2)$, where $IG(\nu/2,\nu s^2/2)$ denotes the inverse 
gamma distribution having mode $s^2$ with $\nu$ being the prior degrees of freedom hyperparameter.

Following Bayes’ theorem, the posterior distribution of $\bfTheta$ is proportional to the product of the likelihood function and the prior 
distribution for $\bfTheta$. The kernel of this distribution does not result in that of a known distribution. We make use of MCMC methods to 
obtain samples from the resulting posterior distribution. In particular, we use the Metropolis-Hastings algorithm. The full conditional 
posterior distributions are described in Appendix A.

\subsection{Model for nominal data}

In this part, we assume the $p$-vector $\bfy_i$ is such that $y_{ij}=k$ is a nominal categorical variable indicating that the $j$-th response variable for sample unit $i$ is in category $k$. The proposed factor model is described as follows:
\begin{align}\begin{array}{rl}\label{nominal_model}
\bfy_i &\sim Categorical \left(\pi_{ij1},\dots,\pi_{ijK}\right),\\
\log\left( \displaystyle\frac{\pi_{ijk}}{\pi_{ij1}}\right) &= \sum_{l=1}^q\beta_{jlk}f_{li}\; ,\, k = 2,\dots,K,
\end{array}
\end{align}
where $\pi_{ijk}$ is the probability that the response variable  lies in category $k$ at component $j$ for unit sample $i$, that is $\sum_{k=1}^K{\pi_{ijk}}=1.$

Note that in this case the response category probability $\pi_{ijk}$ depends on category $k$ through the factor loading $\beta_{jlk}$, that is, 
the latent structure which explains the dependence among the $p$-variables has variant factor loadings, according to each category. As a particular 
case, we can assume the same factor loadings for each category. The binary case is trivially attained by assuming $K=2$ in the model (\ref{nominal_model}).

As well as the ordinal model presented in Section \ref{ordmodel}, the present one suffers from identifiability problems. Since in this case we do not have the latent variable $y_{ij}^*$, the only identifiability restriction needed is due to the invariance of the factor models under orthogonal transformations, which is corrected by constraining the factor loadings matrix to be a block lower triangular matrix with strictly positive diagonal elements. However, as we now have a
factor loadings matrix for each category, this restriction must be imposed to each one of the $K-1$ matrices.

\subsubsection{Inference procedure}

The inference procedure for the model in (\ref{nominal_model}) proceeds in an analogous way to the one described in Subsection \ref{Inf1} for ordered data. Like in Subsection \ref{Inf1}, the likelihood function for $\Theta$ can be written as:
\begin{align*}
l(\bfTheta;{\bf y})&\propto \prod_{i=1}^n\prod_{j=1}^p\prod_{k=1}^K{\pi_{ijk}^{I(y_{ij} = k)}}\\
& = \prod_{i=1}^n\prod_{j=1}^p\prod_{k=2}^{K}{\left( \frac{\exp\left(\sum_{l=1}^q\beta_{jlk}f_{li}\right)}{1+\sum_{m=2}^{K}{\exp\left(\sum_{l=1}^q\beta_{jlm}f_{li}\right)}}\right) ^{I(y_{ij} = k)}}\times \\
& \prod_{i=1}^n\prod_{j=1}^p\left( \frac{1}{1+\sum_{m=2}^{K}{\exp\left(\sum_{l=1}^q\beta_{jlm}f_{li}\right)}}\right)^{I(y_{ij} = 1)} .
\end{align*}

Let us define $\bfbeta^{(k)}$ the $p\times q$-factor loadings matrix for the category $k$. We assign the same prior distribution described in Subsection \ref{Inf1} for the parameter vector $\Theta$. In the same way, we make use of the Metropolis-Hastings algorithm to obtain samples from the resulting posterior distribution. The full conditional posterior
distributions are also described in Appendix A.

\subsection{Choice of the number of factors}\label{sec2.3}

With a specified $q$-factor model, Bayesian analysis using MCMC methods is straightforward. \cite{lopes2004bayesian} treat the case where uncertainty 
about the number of latent factors is assumed in a multivariate factor model. They also discuss reversible jump MCMC and alternative MCMC methods based on bridge sampling.

In the identifiable model, the loadings matrix has
$r = pq - q(q - 1)/2$ free parameters. With $p$ non-zero $\sigma^2_j$ parameters, the
resulting factor form of $\bfDelta$ in (\ref{var_fact_mod}) has $p(q+1)-q(q-1)/2$ parameters, compared with the total $p(p + 1)/2$ in an unconstrained (or $p = q$) model; leading to the
constraint that $p(p + 1)/2 - p(q + 1) + q(q - 1)/2 \geq 0$, which provides at least an upper bound on $q$ \citep{lopes2004bayesian}.

Here, we assume $q$ known and use the previous constraint and polychoric correlation \citep{drasgow2004polychoric} as preliminary exploratory methods to infer the $q$-value. The polychoric correlation is based on the assumption that ordinal polychotomous data can be interpreted as categorization of continuous data, according to some cutoff points. More specifically, it assumes bivariate normality between every pair of variables and estimates the Pearson correlation. 

Another method that can help in elicitation of a value for $q$ is the variance decomposition. It consists basically of a fairly standard to summarize the importance of a common factor by its percentage contribution to the variability of a given attribute. The variance decomposition for variable $j$ is given by: 
\begin{align}\label{DVexpr}
DV_{j} = 100\frac{\sum_{l=1}^{q}\beta_{jl}^2}{\sum_{l=1}^{q} \beta_{jl}^2 + \sigma_j^2} \%. 
\end{align}
Higher values for it enhance the result that the latent factors can explain well a variable regarding their interpretation.

In the applications, we also perform model comparison with different values of $q$ using the Akaike
information criterion (AIC), Bayesian information criterion (BIC) and Widely available information criterion (WAIC). Each of these criteria are described in more detail in Appendix B.

\section{Simulation study}\label{sec3}

In this section we analyze synthetic datasets generated from the proposed models to check our ability to recover the true values of the parameters. 

In the first simulation study we generated several samples from model (\ref{factormodel}) for ordinal data, assuming (\ref{normal_ord}) and (\ref{logist_ord}), and
from model (\ref{nominal_model}) for nominal data, considering $q=1$, in order to examine the performance of the Bayesian estimators under some frequentist measures of the posterior distributions.

In the second simulation study we firstly considered a one and two-factor model for a five-dimensional problem generating from each 
a sample with 300 
observations. Then, we fitted the two models to both datasets, the generating one and the other with different $q$, to evaluate the mismatch 
between the fitted model assumption of $q$ and the data structure
based on a different value of $q$. Then, for several samples generated assuming $q=3$ and $p=9$, we assess model performance by means of model comparison criteria under different values of $q$.

We considered the following hyperparameters in the prior distributions: $C_0=100$, $\nu=0.02$ and $s^2=1$, which yield vague priors.

The MCMC algorithm was implemented in the \texttt{R} programming language, version 3.4.1 \citep{teamR}, in a computer with an Intel(R) Core(TM) 
i5-4590 processor with 3.30 GHz and 8 GB of RAM memory. For each sample and fitted model, we ran two parallel chains starting from very different 
values, letting each chain run for 10,000 iterations, discarded the first 1,000 as burn-in, and stored every 9th iteration to avoid possible autocorrelations within the chains. We used the diagnostic tools available in the CODA package \citep{plummer2006coda} to check convergence of the chains.

\subsection{Bayesian estimators' performance with several samples}\label{study1}

To check the ability of the proposed models for ordinal and nominal data to estimate the parameters, we conducted a Monte Carlo simulation study. For the ordered case, we considered both normal and logistic distributions for the errors, given, respectively in (\ref{normal_ord}) and (\ref{logist_ord}). A total of $200$ samples for each case with $p=5$ variables and assuming $n=300$, $k=4$ and $q=1$ were generated. In each simulation, observations were drawn from a
one-factor model defined by parameters $\bfbeta=(0.99,0.80,0.90,0.70,0.5)'$ and $\bfSigma=diag(0.01,0.05,0.10,0.15,0.20).$ These values were chosen with the aim of generating a dataset with some variables better explained by the common factor than others. The cutoff points were assumed known and determined by quantiles 0.40, 0.75 and 0.90. As measures of performance we computed some frequentist measures of the posterior distributions of the model parameters after reaching convergence.

Table \ref{tab1} reports the root mean square error (RMSE), the mean absolute error (MAE) and the empirical nominal coverage of the 95\% credibility 
intervals measured in percentages (Cov.) for the factor loadings and variances.  Although the coverage of the 
$95\%$ credibility intervals is below the desired level for some of the parameters, the parameters are overall well estimated in both cases. However, the estimation under the probit model seems to be slightly 
better than the logit one, which we believe it is due to characteristics inherent to the models themselves and how the probit and logit functions link the variables to the latent factors. Sometimes standard choices of link
lead to poor fit, and one may instead allow for a choice of the link function
based on the data. A logit link, for instance, is traditionally more robust to extreme linear predictor values, while assuming probit link makes
analytical manipulations easier, as it is based on normality as the factor variables do \citep{albert1993bayesian, congdon2005bayesian}.

\begin{table*}[h!]\caption{ Summary measurements for the point and $95\%$ credibility interval estimates of the parameters of the proposed model 
		for ordinal data in (\ref{factormodel}) with errors having normal (\ref{normal_ord}) and logistic (\ref{logist_ord}) distributions over 200 simulations considering $q=1$. }\vspace{0 cm}
	\begin{center}
		\setlength{\tabcolsep}{4pt}
		\begin{tabular}{c|cccccccccc} \hline
			& $\beta_1$ & $\beta_2$ & $\beta_3$ & $\beta_4$ & $\beta_5$ & $\sigma_1^2$ & $\sigma_2^2$ & $\sigma_3^2$ & $\sigma_4^2$ & $\sigma_5^2$\\\hline
			&\multicolumn{9}{c}{Normal}\\\hline
			RMSE &  0.170 &  0.267 &  0.301 &  0.237 &  0.169 &  0.172 &  0.013 &  0.021 &  0.025 &  0.030 \\ 
  MAE &  0.069 &  0.083 &  0.101 &  0.079 &  0.062 &  0.041 &  0.010 &  0.017 &  0.020 &  0.023 \\ 
  Cov.(\%) & 89.0 & 91.0 & 92.5 & 91.0 & 91.5 & 87.0 & 96.0 & 94.5 & 94.0 & 94.0 \\ 
			\hline
			&\multicolumn{9}{c}{Logistic}\\\hline
			RMSE &  0.207 &  0.329 &  0.364 &  0.288 &  0.208 &  0.244 &  0.013 &  0.024 &  0.025 &  0.030 \\ 
            MAE &  0.088 &  0.107 &  0.123 &  0.095 &  0.075 &  0.053 &  0.010 &  0.019 &  0.020 &  0.024 \\ 
            Cov.(\%) & 88.5 & 89.0 & 88.0 & 90.0 & 92.0 & 95.0 & 94.0 & 92.0 & 93.5 & 94.5 \\
			\hline
		\end{tabular}\label{tab1}
	\end{center}
\end{table*}

\begin{figure}[h!]
	\begin{tabular}{c}
		\hspace{-0.5 cm}
		\subfigure[RMSE]{\includegraphics[scale=0.29]{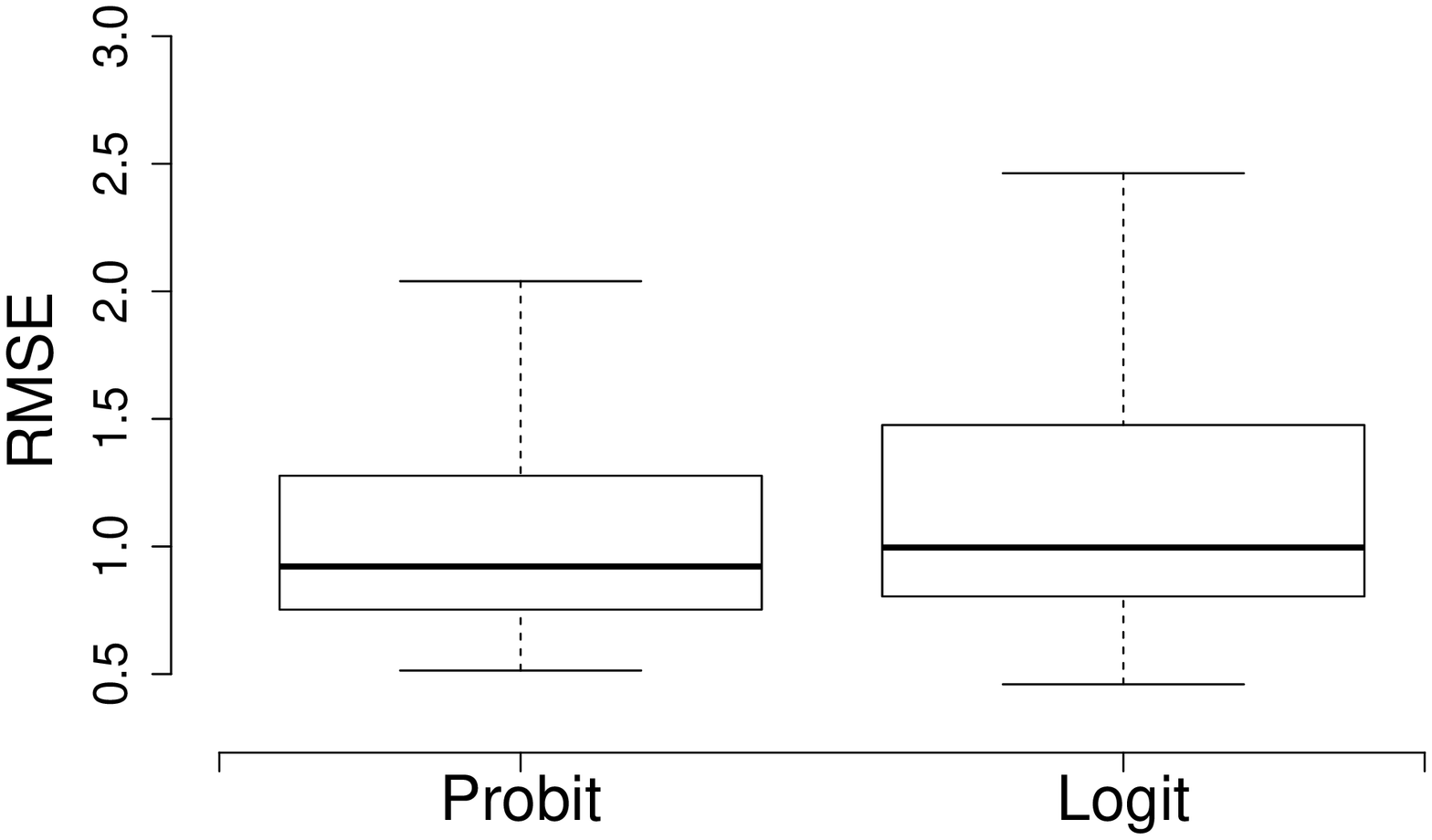}} \hspace{0.00 cm}
		\subfigure[Coverage]{\includegraphics[scale=0.29]{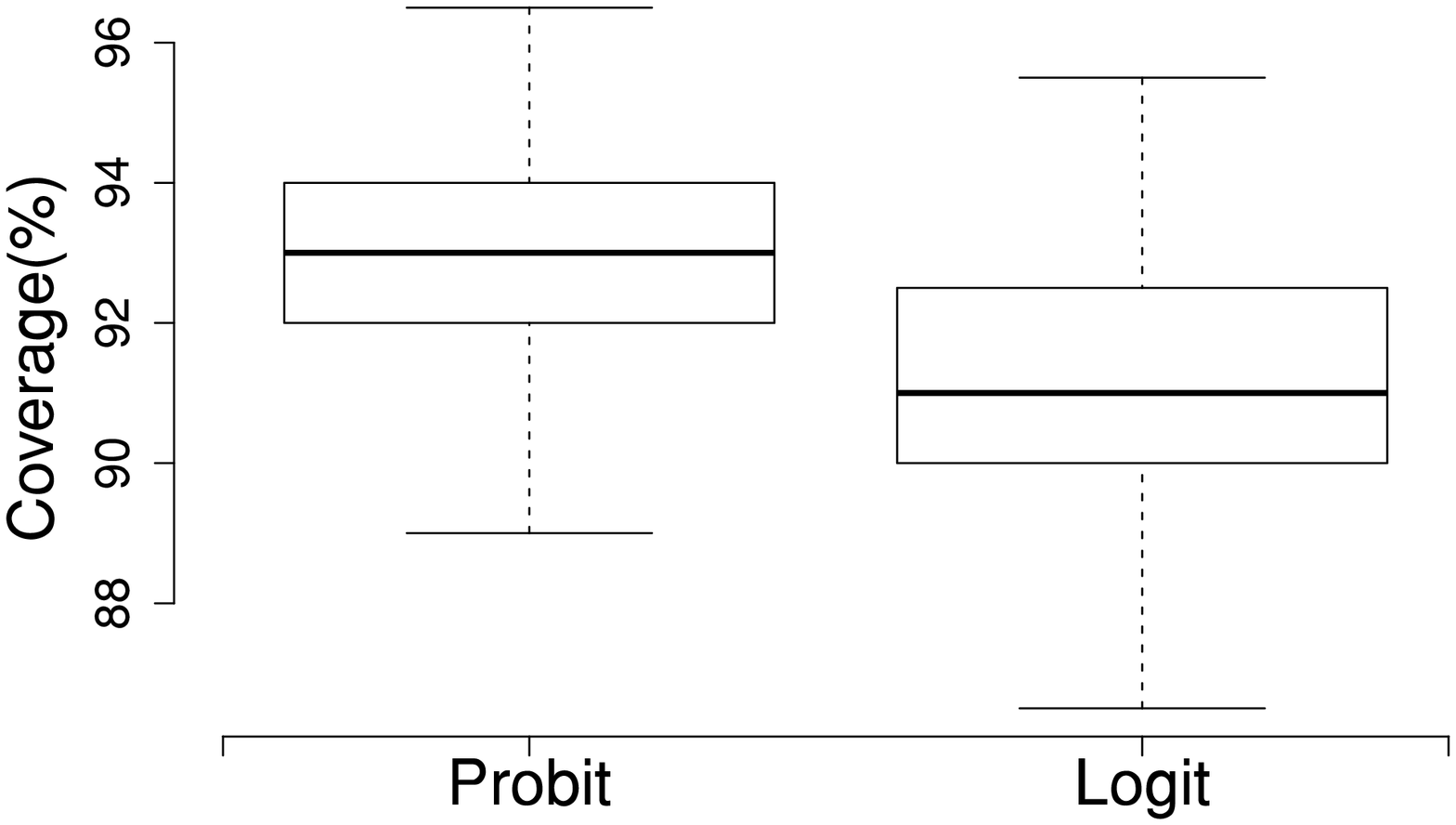}}
	\end{tabular}
	\vspace{-0.1 cm}\caption{ Relative mean absolute error and coverage (\%) for $f_i$'s, $i=1,\dots,300$ and over 200 simulations for the proposed model
		for ordinal data in (\ref{factormodel}) with normal (\ref{normal_ord}) and logistic (\ref{logist_ord}) errors distributed considering $q=1$.}\label{Fig1}
\end{figure}

Figure \ref{Fig1} presents a boxplot with the relative mean square error (RMSE) and coverage of $f_i$'s for all the sample units under all the $200$ replications for each model. Note that, in both cases the latent factors are well estimated, but the probit model also presents smaller RMSE and coverage for the latent factors closer to the nominal level than does logit model.

Then, the same study is done with $200$ samples generated from the model in (\ref{nominal_model}) for nominal data, under similar conditions to 
the previous study on the generation process. Table \ref{tab1a} reports RMSE, MAE and the empirical nominal coverage of the 95\% credibility 
intervals measured in percentages (Cov.) for the factor loadings. In general, the model has a good performance.  Although the coverage of the 
$95\%$ credibility intervals is below the desired level for some parameters, its performance is similar to the results presented in Table \ref{tab1} for the ordinal case. However, the RMSE and MAE are larger than those obtained in the ordinal case. An explanation for that is the number of parameters that increases significantly
for the nominal case, varying by category.

\begin{table*}[h!]\caption{ Summary measurements for the point and $95\%$ credibility interval estimates for parameters of the proposed model for
		nominal data in (\ref{nominal_model}) over 200 simulations considering $q=1$.}\vspace{0 cm}
	\begin{center}
		\setlength{\tabcolsep}{4pt}
		\begin{tabular}{c|ccc|ccc|ccc} \hline
			& \multicolumn{3}{c|}{Category 2} &\multicolumn{3}{c|}{Category 3} &\multicolumn{3}{c}{Category 4} \\\hline
			& RMSE & MAE & Cov.(\%)& RMSE & MAE & Cov.(\%)& RMSE & MAE & Cov.(\%)\\\hline
			$\beta_1$ & 0.367 & 0.280 & 94.5 & 0.345 & 0.341 & 90.6 & 0.543 & 0.261 & 93.0 \\ 
            $\beta_2$ & 0.334 & 0.371 & 93.0 & 0.518 & 0.401 & 90.5 & 0.351 & 0.223 & 91.5 \\ 
            $\beta_3$ & 0.308 & 0.254 & 90.0 & 0.506 & 0.350 & 89.0 & 0.357 & 0.111 & 94.0 \\ 
            $\beta_4$ & 0.527 & 0.348 & 90.0 & 0.346 & 1.346 & 95.0 & 0.347 & 0.305 & 94.5 \\ 
            $\beta_5$ & 0.503 & 0.174 & 89.1 & 0.541 & 0.175 & 94.0 & 0.473 & 0.328 & 92.5 \\  \hline
		\end{tabular}\label{tab1a}
	\end{center}
\end{table*}

\begin{figure}[h!]
	\begin{tabular}{c}
		\hspace{-0.5 cm}
		\subfigure[RMSE]{\includegraphics[scale=0.27]{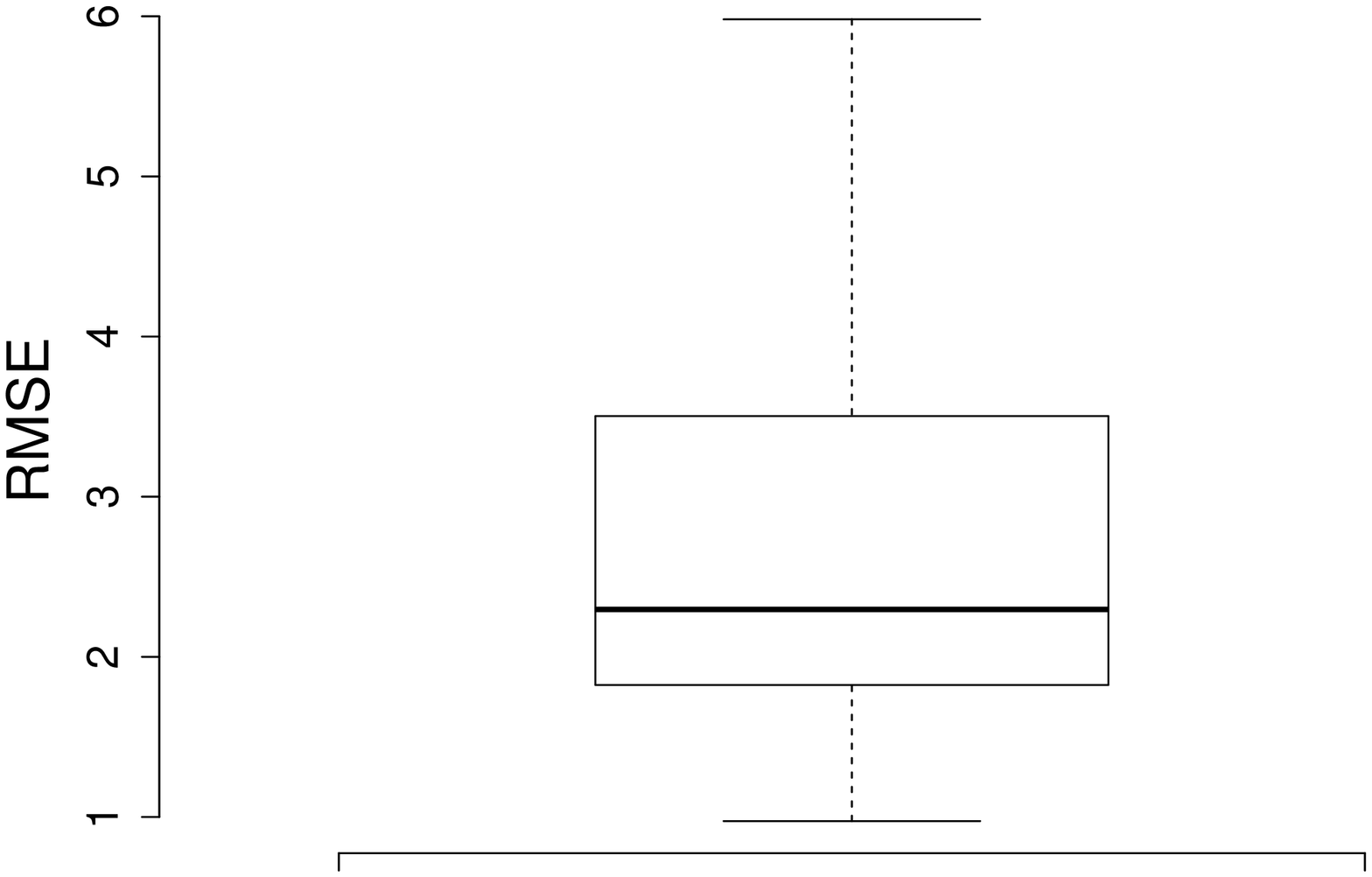}} \hspace{-0.1 cm}
		\subfigure[Coverage]{\includegraphics[scale=0.27]{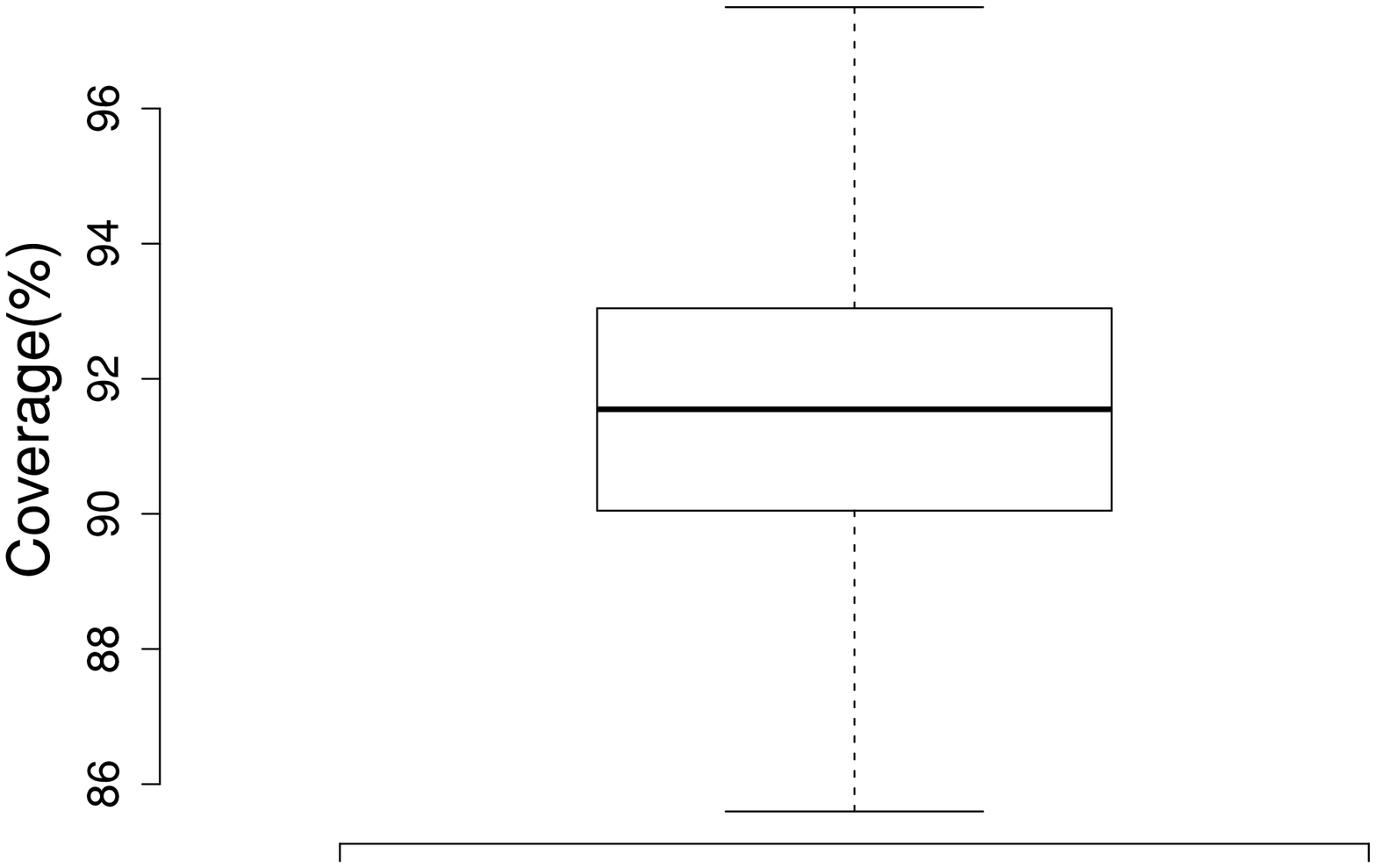}}
	\end{tabular}
	\vspace{-0.1 cm}\caption{Relative mean absolute error and coverage (\%) for $f_i$'s, $i=1,\dots,300$ and over 200 simulations of the proposed model
		for nominal data in (\ref{nominal_model}) considering $q=1$.}\label{Fig1a}
\end{figure}

Finally, Figure \ref{Fig1a} shows a boxplot with RMSE and coverage (\%) of $f_i$'s for all the sample units under all the 200 replications of the model in (\ref{nominal_model}). The latent factors are also well estimated in this case.

In general, although the results obtained in this section point to a good estimation of the parameters of the proposed models, we conclude that the probit model in (\ref{logist_ord}) for the ordered categorical variables performs better under the frequentist properties considered in the simulation study.

\subsection{Comparing different values of $q$}\label{study2}
The aim of this study is to explore the effect on results of fixing a value for $q$ different from those used in the generation of the data. 
In particular, we just considered the proposed model in (\ref{factormodel}) with normal error in (\ref{normal_ord}), 
because the conclusions are the same for this one and the models in (\ref{logist_ord}) and (\ref{nominal_model}). 

\subsubsection{Study with one sample}

\begin{figure}[h!]
	\centering
	{\includegraphics[scale=0.55]{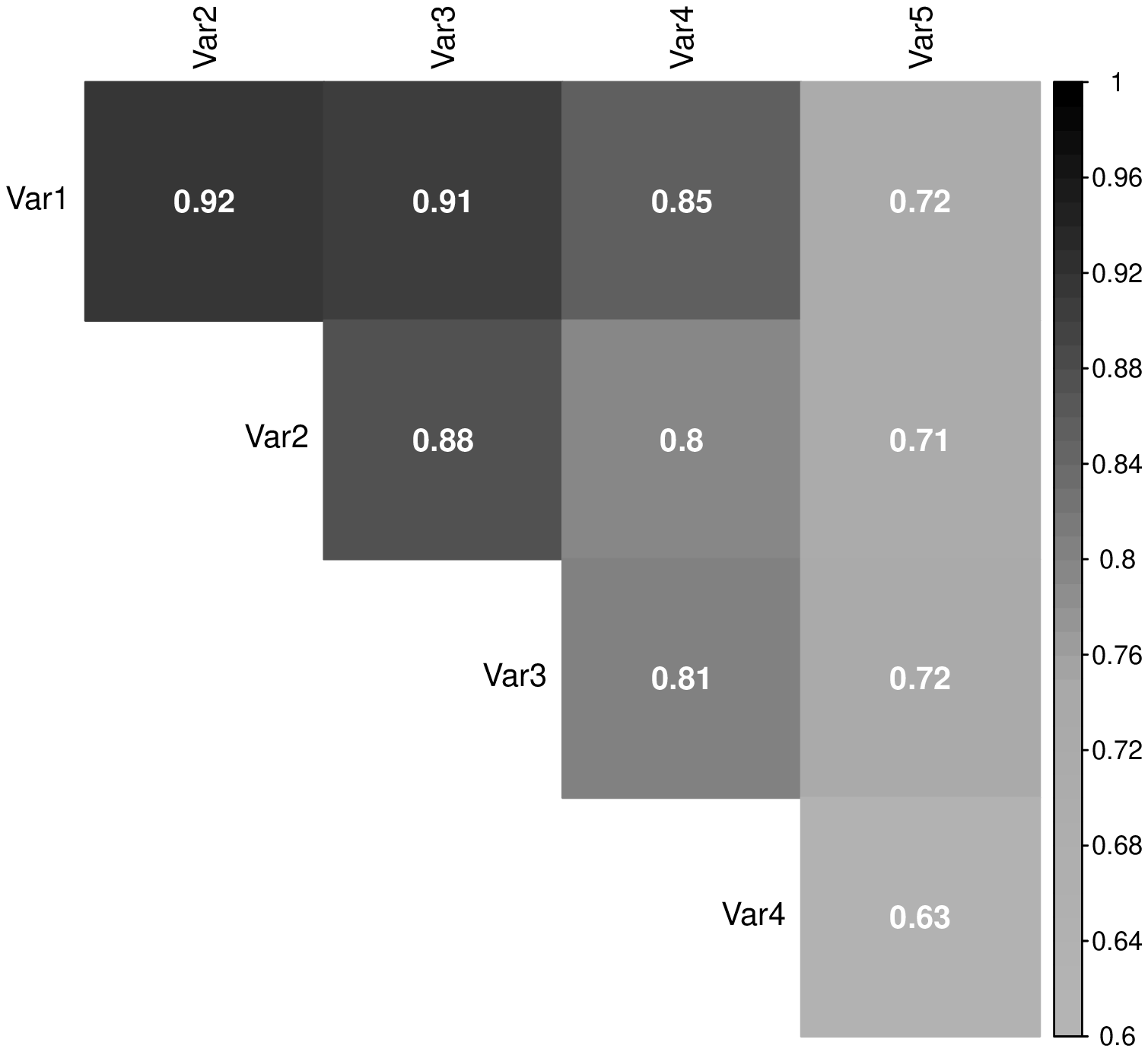}}
	\vspace{-0.3 cm} \caption{Polychoric correlation matrix calculated for the artificial dataset generated by the probit model in (\ref{normal_ord}) assuming $q=1$ factor.}\label{fig_polich1}
\end{figure}

In this initial study, we just generated one sample from the one-factor for the five-dimensional variable, with the same parameter 
values described in Subsection \ref{study1}. Figure \ref{fig_polich1} presents the polychoric correlation for the dataset generated. 
Note that, as expected, due to the factor loading structure, all the components present strong correlation, with the last one being less 
correlated with the other components.

\begin{figure}[h!]
	\centering
	{\includegraphics[scale=0.55 ]{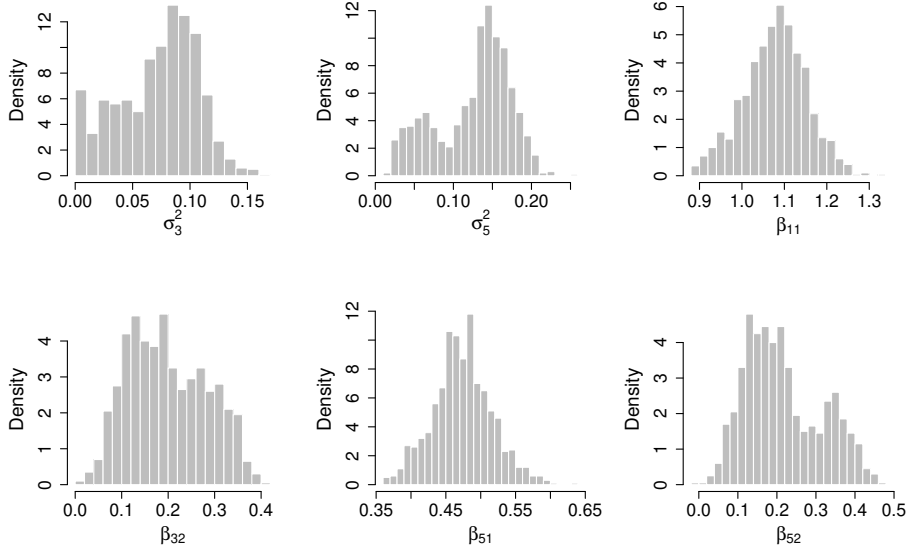}}
	\vspace{-0.3 cm} \caption{Marginal posteriors of $\sigma^2_3$, $\sigma^2_5$, $\beta_{11}$, $\beta_{32}$, $\beta_{51}$ and $\beta_{52}$ from analysis of the simulated dataset from a one-factor structure but analyzed using a model
		with $q = 2$ factors.}\label{fig2}
\end{figure}

We fitted the same model, but assuming $q=1$ and then $q=2$. For $q=1$, as expected from the study in Subsection \ref{study1}, the convergence was achieved and the parameters were well-recovered. On the other hand, Figure \ref{fig2} displays marginal posterior densities of some of the idiosyncratic variances and factor loadings from the analysis assuming $q=2$. It is possible to observe multimodality in both parameters induced
by multiple local maxima in the likelihood functions when the specified value of $q$ is larger than consistent with the data. This arises due to the mismatch between the model assumption of $q=2$ and the data structure based on $q=1$.  Encountering such multimodality in posterior samples from a specified model can therefore be taken as an indication that
the chosen value of $q$ is too large. This particular behavior happens similarly to the standard normal factor model \citep{lopes2004bayesian}.

Then, we performed model comparison for one- and two-factor models using the following criteria: AIC, BIC and WAIC. See Appendix B
for details. Table \ref{tab41} shows the different model comparison criteria computed for assessment of the number of factors. 
The factor model fitted assuming $q=1$ provides the lowest values under all criteria. Thus, all the different criteria, 
together with the multimodality behavior previously described, suggest that $q=1$ performs best between the two models considered. Among the three 
criteria considered, BIC was the one which showed the biggest difference between the one- and two-factor models.

\begin{table*}[h!]\caption{ Model comparison criteria with $q=1$ and $q=2$ for the dataset generated assuming $q=1$. Under both criteria, the smallest values (in italics) indicate the best model among those fitted.}\vspace{0 cm}
	\begin{center}
		\begin{tabular}{c|cccccccccc} \hline
			& $q=1$ & $q=2$\\\hline
			AIC & {\it 2390.585} &  2392.176\\
			BIC & {\it 2437.623} & 2458.029 \\
			WAIC & {\it 2398.977} & 2400.306 \\\hline
		\end{tabular}\label{tab41}
	\end{center}
\end{table*}

Moreover, Table \ref{tab2} reports the posterior mean and 95\% credibility interval of the variance decomposition, given by (\ref{DVexpr}), for each 
variable. Observe that they are well-estimated when the model that generated the data is fitted to the data, while the model assuming $q=2$ presents, in general, higher values for $DV_j$. This happens because increasing the number of factors decreases the residual variability, as expected.

\begin{table*}[h!]\caption{ Posterior mean and $95\%$ credibility interval in parentheses for $DV_j$, $j=1,\dots,5$, given by expression (\ref{DVexpr}), for the model that generated the dataset ($q=1$) and the model assuming $q=2$.}\vspace{0 cm}
	\begin{center}
		\begin{tabular}{c|ccccccccc} \hline
			& True & $q=1$ & $q=2$\\\hline
			$DV_1$ &  99.0 & 98.4 (95.4,99.9) & 99.6 (98.1,100)\\\hline
			$DV_2$ & 92.8 & 91.2 (86.5,94.7) & 90.7 (86.1,94.2)\\\hline
			$DV_3$ & 89.0 & 89.5 (84.8,93.3) & 92.4 (86.2,99.6)\\\hline
			$DV_4$ & 76.6 & 75.8 (67.7,82.3) & 75.7 (67.8,82.4)\\\hline
			$DV_5$ & 55.6 & 58.6 (48.7,67.7) & 68.7 (50.2,92.8)\\\hline
		\end{tabular}\label{tab2}
	\end{center}
\end{table*}

On the other hand we repeated the previous study, now generating a sample from the same proposed factor model but assuming $q=2$, and fixing in this 
case $\bfbeta' = \left(
\begin{array}{ccccc}
0.99 & 0.00 & 0.90 & 0.00 & 0.50\\
0.00 & 0.99 & 0.00 & 0.90 & 0.50
\end{array}\right).$

\begin{figure}[h!]
	\centering
	{\includegraphics[scale=0.55]{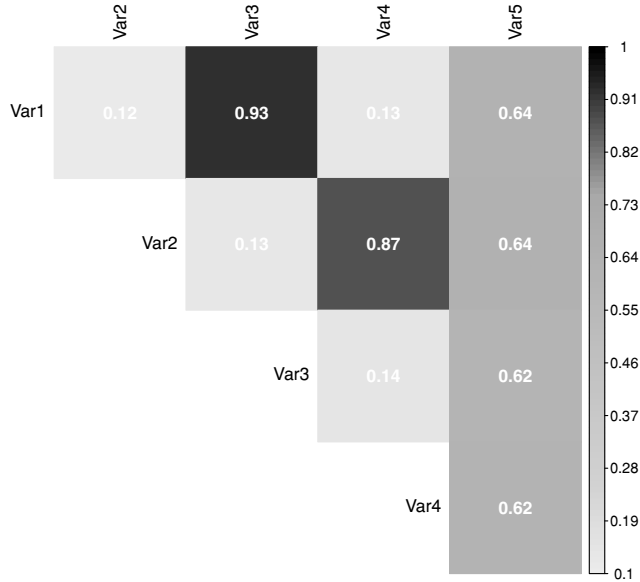}}
	\vspace{-0.3 cm} \caption{Polychoric correlation matrix calculated for the artificial dataset generated by the probit model in (\ref{normal_ord}) assuming $q=2$ factors.}\label{fig_polich2}
\end{figure}

Figure \ref{fig_polich2} presents the polychoric correlation calculated for the dataset generated assuming $q=2$. Note that, as expected, due to the factor loading structure, variables 1 and 3 are connected to the first factor, while variables 2 and 4 are connected to the second factor, thus presenting a strong correlation between them. On the other hand, the 5th component is correlated with all the other components.

In this case, if we fit a factor model with $q=1$ for a sample generated from a two-factor model, as can be seen in Table \ref{tab3}, a significant decrease happens in the $DV_j$ values for the variables $j=2$ and 4, which are just associated with the second factor. In particular, variable $j=5$ is the only one associated with the second factor, which presents just a partial decrease in $DV$, because it is also explained by the first factor, as can be seen in the $\bfbeta$ formulation.

\begin{table*}[h!]\caption{ Posterior mean and $95\%$ credibility interval in parentheses for $DV_j$, $j=1,\dots,5$, given by expression (\ref{DVexpr}), for the model that generated the dataset ($q=2$) and the model assuming $q=1$.}\vspace{0 cm}
	\begin{center}
		\begin{tabular}{c|ccccccccc} \hline
			& True & $q=1$ & $q=2$\\\hline
			$DV_1$ & 99.0 & 96.5 (90.7,99.8) & 97.7 (91.8,99.9)\\\hline
			$DV_2$ & 95.1 & 5.1 (0.7,12.0) & 97.3 (89.4,99.9)\\\hline
			$DV_3$ & 89.0 & 94.8 (88.9,99.5) & 93.7 (89.1,99.2)\\\hline
			$DV_4$ & 84.4 & 5.6 (0.7,13.0) & 84.3 (76.5,93.1)\\\hline
			$DV_5$ & 71.4 & 45.0 (34.0,56.3) & 77.1 (69.5,84.0)\\\hline
		\end{tabular}\label{tab3}
	\end{center}
\end{table*}

Unlike the previous study, in which the convergence was not achieved, as shown in Figure \ref{fig2}, and the multimodalities clearly suggest 
that $q=2$ is large for a dataset generated from a $q=1$ factor model, in this case, since convergence was achieved for both cases, 
besides $DV$ values, we strongly suggest model comparison criteria to be used in choosing $q-$ value. Table \ref{tab4} shows the different model comparison 
criteria computed for 
assessment of the number of
factors. All the different criteria suggest that $q=2$ performed best among the fitted ones. Different from the results in Table \ref{tab41}, in this case 
we observe a bigger difference between $q=1$ and $q=2$ models fit for all the criteria.

\begin{table*}[h!]\caption{
		Model comparison criteria with $q=1$ and $q=2$ for the dataset generated assuming $q=2$. Under both criteria, the smallest values (in italics) 
		indicate the best model among those fitted.}\vspace{0 cm}
	\begin{center}
		\begin{tabular}{c|cccccccccc} \hline
			& $q=1$ & $q=2$\\\hline
			AIC & 3296.399 & {\it 2954.980}\\
			BIC & 3343.437 & {\it 3020.833}\\
			WAIC & 3307.625 & {\it 2969.737} \\\hline
		\end{tabular}\label{tab4}
	\end{center}
\end{table*}

However, as pointed by \cite{lopes2004bayesian}, there might be some uncertainty in the results presented by the model selection criteria. 
Some of them often tend to prefer models with higher number of factors. We investigate this statement for the proposed model in the following study.

\subsubsection{Study with several samples}
This second analysis is done in the same fashion as the previous study, but now using 50 datasets simulated from the ordinal model in (\ref{factormodel}) assuming normal distribution for the error, as described in equation (\ref{normal_ord}), with $p=9$ variables 
and $q = 3$ 
factors. The generating model in this case has parameters
\begin{align*}
\bfbeta' = & \left(
\begin{array}{ccccccccccccccc}
0.99 & 0.00 & 0.00 & 0.99 & 0.99 & 0.00 & 0.00 & 0.00 & 0.00\\
0.00 & 0.95 & 0.00 & 0.00 & 0.00 & 0.95 & 0.95 & 0.00 & 0.00\\
0.00 & 0.00 & 0.90 & 0.00 & 0.00 & 0.00 & 0.00 & 0.90 & 0.90\\
\end{array}\right), \\
\Sigma = & diag(0.02,0.19,0.36,0.02,0.02,0.19,0.19,0.36,0.36).
\end{align*}

The aim of this study is to assess model performance by means of comparison 
criteria in this new context. To this end, we fitted models with 
$q=1, 2, 3, 4$ and $5$ to the several samples and computed for each one the values of the comparison criteria. The value $q=5$ was the maximum to be tested, respecting the upper bound on $q$ generated by the constrained imposed to become the model identifiable, as described in Subsection \ref{sec2.3}. Table \ref{tab9} presents the proportion of times, in percentage, that each $q$-factor model achieved the smallest value of each comparison criterion. All 
the different criteria suggest, in most of the times, that the model with $q=3$ outperformed the other ones. Samples for which the model with $q=3$ was not pointed by the 
comparison criteria, in general, pointed the model with $q=4$ as the best one. Models whose number of factors are smaller than 3 were never selected. Moreover, the model with $q=5$ factors was almost never selected, which also suggests that a model with a larger number of factors could not fit the data well. Among the considered comparison criteria, we notice that BIC is the one that chose the true generating model most of time, besides being easier to compute and demanding less computational effort when compared to WAIC, for example.

\begin{table*}[h!]\caption{ Percentage of samples which indicates $q-$ factor model as the best for each comparison criteria among the 
		50 samples generated
		assuming $q=3$.}\vspace{0 cm}
	\begin{center}
		\begin{tabular}{c|cccccccccc} \hline
			& $q=1$ & $q=2$ & $q=3$ & $q=4$ & $q=5$\\\hline
			AIC  & 0 & 0 & 72 & 26 & 2\\
			BIC  & 0 & 0 & 82 & 18 & 0\\
			WAIC & 0 & 0 & 80 & 20 & 0\\\hline
		\end{tabular}\label{tab9}
	\end{center}
\end{table*}

In addition to the comparison criteria results, multimodality behavior in posterior samples as well as the computation of the values of $DV$ may be explored in this case to help choosing the number of factors.

\section{Data analysis}\label{sec4}

In order to illustrate the method proposed, we explore the factor structure underlying the emotions dataset collected by the Motivational State Questionnaire (MSQ). It was developed to study emotions in laboratory and field settings. The data are available in the package \texttt{psych} \citep{psych} of the software R with the name \texttt{msq}, and can be well described in terms of a two dimensional solution of energy versus tiredness and tension versus calmness. We worked here with a random sample of $n=500$ observations selected from the whole dataset.

The MSQ is composed of 72 items, which represent the full affective space: 20 items taken from the Activation-Deactivation Adjective Check List \citep{thayer1986activation}, and 18 from the Positive and Negative Affect Schedule \citep{watson1988development} along with the items used by \citep{larsen1992promises}. The response format is a four-point scale that asks the respondents to indicate their current standing with the following rating scale: ``Not at all", ``A little", ``Moderately" and ``Very much". 

For this application we just considered the following 10 items extracted from \cite{thayer1986activation}:  ``sad", ``blue", ``unhappy", ``gloomy", ``grouchy", ``jittery", ``anxious", ``nervous", ``fearful", and ``distressed". The cuttoff points were previously estimated by the inverse standard normal distribution function evaluated at the cumulative marginal proportions, as described by \cite{drasgow2004polychoric}.

\begin{figure}[h!]
	\hspace{-0.8 cm}\centering
	{\includegraphics[scale=0.6]{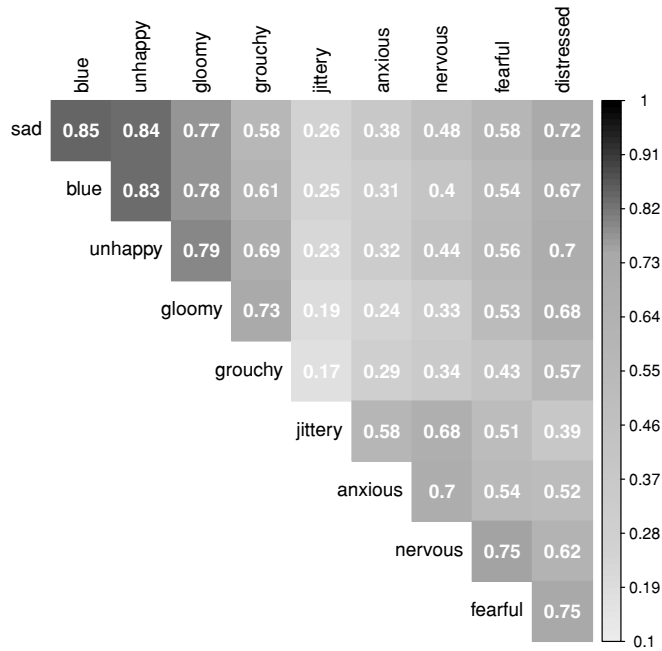}}
	\vspace{-0.3 cm} \caption{Polychoric correlation matrix calculated for the MSQ dataset.}\label{fig_polich_appl}
\end{figure}

In an illustrative analysis we calculated the polychoric correlation (see Figure \ref{fig_polich_appl}) to explore the correlation structure 
underlying the dataset and the uncertainty about the number of latent factors. Note that the variables ``sad", ``blue", ``unhappy" and ``gloomy" 
seem to be mutually correlated, being well explained by one factor, while ``jittery", ``anxious" and ``nervous" can be explained by another 
independent latent factor. Moreover, ``distressed", although well explained by the first factor, is the one in the first group that is most strongly 
correlated with the second factor. In this way, beyond the intrinsic idea behind the behaviors explained by the variables, the polychoric correlation seems to provide a preliminary suggestion that a factor model with 2 factors is reasonable for this case.

Thus, in this analysis we fitted the proposed model in (\ref{factormodel}) assuming normal error in (\ref{normal_ord}), $q=1, 2$ and $3$ and the same prior distribution specified in the simulated examples in Section \ref{sec3}. For each model, we ran two parallel chains starting from very different values, and let each chain run for 20,000 iterations, discarded the first 1,000 as burn-in, and stored every 19th sample value. We also used the diagnostic tools available in the package CODA \citep{plummer2006coda} to check convergence of the chains. 
The marginal posteriors for $q=1$ and $q=2$ are all unimodal, while multiple modes appear in the analysis of a three-factor model. Figures 
\ref{fig3}, \ref{fig4}, \ref{fig5}, \ref{fig6} and \ref{fig7} in 
Appendix C present histograms with the marginal posteriors of the factor loadings and idiosyncratic variances when fitting a one-, two- and three-factor structure. This fact provides additional support that a $q=3$ is too large in this case.

 Although the exploratory analysis indicates $q=2$ a reasonable choice for the dataset, the maximum value of $q$ to be tested satisfying the restriction presented in Subsection \ref{sec2.3} is $q=6$. Thus, we also fitted the proposed model for $q=4, 5$ and $6$, however, we have some problems in reaching the convergence, which is usual when the specified value of $q$ is larger  than  consistent  with the  data. This is due to the  mismatch  between  the model assumption and the data structure based in a smaller value of $q$, as discussed in the studies presented in Section \ref{study2}.

Additionally, Table \ref{tab5} shows the different model comparison criteria computed for assessment of the number of factors. The model fitted 
assuming $q=1$ provides the highest values under all criteria. While AIC and BIC point to the two-factor model as the best, WAIC points
to $q=3$, although its value is close to the one obtained with $q=2$.
In particular, all the different criteria, together with the multimodality behavior previously described, suggest that $q=2$ performs best among 
the fitted models.

\begin{table*}[h!]\caption{Model comparison criteria under $q=1$, $2$ and $3$ probit factor model for the MSQ dataset. Under all 
		three criteria, the smallest values (in italics) indicate the best model among those fitted.}\vspace{0 cm}
	\begin{center}
		\begin{tabular}{c|cccccccccc} \hline
			& $q=1$ & $q=2$ & $q=3$\\\hline
			AIC  & 7348.685 & {\it7056.136} & 7100.842\\
			BIC & 7452.977 & {\it7207.36} & 7293.783\\
			WAIC & 7371.821 & 7098.035 &  {\it7093.390}\\\hline
		\end{tabular}\label{tab5}
	\end{center}
\end{table*}

In particular, from the MCMC analysis of the $q = 2$ factor model, we report in Table \ref{tab7} the loadings' posterior means with associated 95\% credibility intervals. All the
loadings, except $\beta_{32}$ and $\beta_{52}$, associated with the second factor, are significant. In general, we can observe that the items concerning sadness and depression present high positive loadings related to the first factor, whereas the items measuring anxiety and tension have high negative loadings in correspondence of the second factor. In the group of variables well explained by the first factor, ``fearful" and ``distressed" are the ones with the highest factor loadings associated with the second factor, which is expected due to the nature of this feeling. All these results are in accordance with the polychoric correlation presented in Figure \ref{fig_polich_appl}.

\begin{table*}[h!]\caption{ Posterior summary, means and 95\% posterior credible intervals of the factor loadings in the MSQ dataset.}\vspace{0 cm}
	\begin{center}
		\begin{tabular}{c|c|cc||c|cc} \hline
			Variable & Loading & Mean & 95\% CI & Loading  & Mean & 95\% CI \\\hline
			sad & $\beta_{11}$  & 0.96 &(0.86,1.08) & & & \\
			blue & $\beta_{21}$  & 0.94 &(0.83,1.06) & $\beta_{22}$  & 0.09 & (0.01,0.20) \\
			unhappy & $\beta_{31}$  & 0.97 &(0.86,1.08) & $\beta_{32}$  & 0.08 &(-0.01,0.18) \\
			gloomy & $\beta_{41}$  & 0.90 &(0.80,1.03) & $\beta_{42}$  & 0.19 &(0.09,0.30) \\
			grouchy & $\beta_{51}$  & 0.76 &(0.65,0.89) & $\beta_{52}$  & 0.10 &(-0.01,0.21) \\
			jitery & $\beta_{61}$  & 0.31 &(0.19,0.44) & $\beta_{62}$  & -0.68 &(-0.81,-0.56) \\
			anxious & $\beta_{71}$  & 0.42 &(0.30,0.54) & $\beta_{72}$  & -0.67 &(-0.80,-0.54) \\
			nervous & $\beta_{81}$  & 0.55 &(0.42,0.69) & $\beta_{82}$  & -0.83 &(-0.97,-0.71) \\
			fearful & $\beta_{91}$  & 0.71 &(0.56,0.84) & $\beta_{92}$  & -0.53 &(-0.66,-0.39) \\
			distressed & $\beta_{10,1}$  & 0.84 &(0.72,0.96) & $\beta_{10,2}$  & -0.28 &(-0.40,-0.18) \\\hline
		\end{tabular}\label{tab7}
	\end{center}
\end{table*}

Table \ref{tab8} presents the posterior summary, means and 95\% credible intervals, of the idiosyncratic variances for the MSQ dataset, 
for the probit $q=2$ factor model fitted. The variables ``grouchy", ``jittery" and ``anxious" present highest estimated variance.

\begin{table*}[h!]\caption{Posterior summary, means and 95\% credible intervals, of the idiosyncratic variances for the MSQ dataset.}\vspace{0 cm}
	\begin{center}
		\begin{tabular}{c|cc} \hline
			& Mean & 95\% CI \\\hline
			$\sigma^2_{1}$  & 0.17 &(0.12,0.23)\\
			$\sigma^2_{2}$  & 0.19 &(0.13,0.25) \\
			$\sigma^2_{3}$  & 0.16 &(0.11,0.22) \\
			$\sigma^2_{4}$  & 0.23 &(0.16,0.31) \\
			$\sigma^2_{5}$  & 0.49 &(0.39,0.60) \\
			$\sigma^2_{6}$  & 0.49 &(0.37,0.63) \\
			$\sigma^2_{7}$  & 0.45 &(0.33,0.58) \\
			$\sigma^2_{8}$  & 0.09 &(0.00,0.20) \\
			$\sigma^2_{9}$  & 0.29 &(0.20,0.40) \\
			$\sigma^2_{10}$ & 0.31 &(0.23,0.39) \\\hline
		\end{tabular}\label{tab8}
	\end{center}
\end{table*}

Finally, Table \ref{tab6} presents the percentage of the variance of each variable explained by the two-factor model, given by expression in (\ref{DVexpr}), but now separated for each factor. In general variables associated with sadness (first factor) have higher percentage of 
the variance explained by the latent factors. As expected by Table \ref{tab8}, the variables ``grouchy", ``jittery" and ``anxious" present smaller percentage of variance explained by the factor model. A $q=3$ factor
model would move most of this variability over to the third factor. However, as previously discussed, multiple modes appear in its analysis.

\begin{table*}[h!]\caption{ {\it Percentage of the variance of each variable in (\ref{DVexpr}) explained by each factor
			in analysis of the MSQ dataset assuming the probit two-factor model.}}\vspace{0 cm}
	\begin{center}
		\begin{tabular}{c|cccccccccc} \hline
			& $DV_1$ & $DV_2$ & $DV_3$ & $DV_4$ & $DV_5$ & $DV_6$ & $DV_7$ & $DV_8$ & $DV_9$ & $DV_{10}$\\\hline
			Total & 84.6  & 82.5 & 86.0 & 78.9 & 54.8 & 53.5 & 58.5 & 91.5 & 72.9 & 72.1\\
			factor 1 & 84.6 & 81.5 & 85.2 & 75.3 & 53.7 & 9.6 & 16.5 & 28.0 & 46.4 & 64.5\\
			factor 2 & 0.0 & 1.0 & 0.8 & 3.7 & 1.2 & 43.9 & 42.0 & 63.4 & 26.5 & 7.6\\\hline
		\end{tabular}\label{tab6}
	\end{center}
\end{table*}

\section{Conclusion}\label{sec5}

We propose a factor model for ordered and non-ordered correlated polychotomous datasets that can be explained by a smaller number of latent factors. 
When categories are ordered the proposed model assumes that the observed category is related  to  an  underlying  latent  continuous  variable, 
which  is  modeled  according to a Gaussian or logistic latent process, centered on the non-observable reduced factors structure. This is 
equivalent to assuming the probability in a category a probit or a logit link function, respectively. On the other hand, when the categories 
do not present an order, the categorical logistic model is assumed for the vector of observed categories. The model presented assumes the factor loadings vary according to the category, but particular cases can be trivially attained.

Inference is performed under the Bayesian paradigm. A sample from the posterior distribution can be obtained
using MCMC methods. In particular, we use the Metropolis-Hasting algorithm.

Simulation studies in Section \ref{sec3} show that we are able to recover the true values of the parameters used to generate the data. 
Furthermore, aspects like multimodalities, variance decomposition and appropriate analysis of some model comparison criteria, 
as in the normal factor model, can be used in this case to help in choosing
the number of latent factors.

We also analyze a subset of the MSQ dataset. The results obtained from the model fit seem to be promising, as they are in line with the results obtained using polychoric correlation, where there is is a very clear two-factor structure of energetic and tense arousal. However, an advantage of the proposal is its flexibility, since it is a statistical model that estimates the latent factors.

\section*{Funding}
Capdeville was supported by a scholarship from Funda\c{c}\~ao de Amparo \`a Pesquisa do Estado do Rio de Janeiro (FAPERJ).

\appendix
\section{Full conditional posterior distributions of the parameters in the proposed models}\label{Ap1}

We used block MCMC to improve the computational efficiency of the estimation procedure of $\bfbeta$ and $\bff$. For the ordinal model in (\ref{factormodel}),
let $\bfbeta_j = (\beta_{j1},...,\beta_{jq})'$ and $\bff_i = (f_{1i},...,f_{qi})'$. For the prior distribution defined in Subsection \ref{Inf1}, the full conditional distributions are given by:
\begin{align*}
p(\bfbeta_j|\bfSigma,\bff,\bfy) & \propto \prod_{i=1}^n \prod_{k=1}^K {\pi_{ijk}^{I(y_{ij}=k)}} \prod_{l=1}^q \exp{\left(\frac{-1}{2C_0}\beta_{jl}^2\right)}[I(j>l) + \\ & I(\beta_{jl}>0)I(j=l) + I(\beta_{jl}=0)I(j<l)] \\
p(\sigma^2_j|\bfbeta,\bff,\bfy) & \propto \prod_{i=1}^n \prod_{k=1}^K {\pi_{ijk}^{I(y_{ij}=k)}}(\sigma_j^{-2})^{\nu/2 + 1}\exp\left(-\frac{1}{2}\nu s^2\sigma_j^{-2}\right)\\
p(\bff_i|\bfbeta,\bfSigma,\bfy) & \propto \prod_{j=1}^p \prod_{k=1}^K {\pi_{ijk}^{I(y_{ij}=k)}}\exp\left(\frac{-\bff_i'\bff_i}{2} \right),
\end{align*}
where:
\begin{align*}
{\pi_{ijk}} & = {\left[\Phi\left(\displaystyle\frac{\alpha_{j,k} - \sum_{l=1}^q \beta_{jl}f_{li}}{\sigma_j}\right) - \Phi\left(\displaystyle\frac{\alpha_{j,k-1} - \sum_{l=1}^q \beta_{jl}f_{li}}{\sigma_j}\right)\right]}
\end{align*}
for the probit link function, or:
\begin{align*}
{\pi_{ijk}} & = \frac{\exp\left(\displaystyle\frac{\alpha_{j,k} - \sum_{l=1}^q \beta_{jl}f_{li}}{\sigma_j}\right)}{1+\exp\left(\displaystyle\frac{\alpha_{j,k} - \sum_{l=1}^q \beta_{jl}f_{li}}{\sigma_j}\right)} - \frac{\exp\left(\displaystyle\frac{\alpha_{j,k-1} - \sum_{l=1}^q \beta_{jl}f_{li}}{\sigma_j}\right)}{1+\exp\left(\displaystyle\frac{\alpha_{j,k-1} - \sum_{l=1}^q \beta_{jl}f_{li}}{\sigma_j}\right)}
\end{align*}
for the logit link function.
The full conditional distributions do not have a closed form. We make use of  the Metropolis-Hastings algorithm with a random walk proposal distribution to obtain samples from each of them.

For the nominal model in (\ref{nominal_model}), let $\bfbeta_{jk} = (\beta_{j1k},...,\beta_{jqk})'$ and $\bff_i = (f_{1i},...,f_{qi})'$. The full conditional distributions are given by:
\begin{align*}
p(\beta_{jk}|\bff,\bfy) & \propto \prod_{i=1}^n {\pi_{ijk}^{I(y_{ij}=k)}} \prod_{l=1}^q \exp{\left(\frac{-1}{2C_0}\beta_{jlk}^2\right)}[I(j>l) + \\
& I(\beta_{jlk}>0)I(j=l) +  I(\beta_{jlk}=0)I(j<l)] \\
p(\bff_i|\bfbeta,\bfy) & \propto \prod_{j=1}^p \prod_{k=1}^K {\pi_{ijk}^{I(y_{ij}=k)}}\exp\left(\frac{-\bff_i'\bff_i}{2} \right),
\end{align*}
where:
\begin{align*}
{\pi_{ijk}} & = \frac{1}{1+\sum_{m=2}^{K}{\exp\left(\sum_{l=1}^q\beta_{jlm}f_{il}\right)}},\mbox{ for } k = 1\\
{\pi_{ijk}} & = {\frac{\exp\left(\sum_{l=1}^q\beta_{jlk}f_{il}\right)}{1+\sum_{m=2}^{K}{\exp\left(\sum_{l=1}^q\beta_{jlm}f_{il}\right)}}},\mbox{ for } k = 2,\dots,K. 
\end{align*}
Also, the full conditional distributions do not have a closed form. We make use of  the Metropolis-Hastings algorithm with a random walk proposal distribution to obtain samples from each of them.

\section{Model comparison criteria} \label{sec:comparison}

In this section we briefly describe the model comparison criteria used to compare the fitted models in Sections \ref{sec3} and \ref{sec4}.

\subsection{Akaike and Bayesian information criteria} 
The Akaike information criterion (AIC) and Bayesian information criterion (BIC) are defined, respectively as:
\begin{align*}
\rm AIC & = -2\log\, p(\bfy\mid q,\hat{\bfbeta},\hat{\bfSigma})+2m,\\
\rm BIC & = -2\log\,p(\bfy\mid q,\hat{\bfbeta},\hat{\bfSigma})+\log(n)m,
\end{align*}
where $m=p(q+1)-q(q-1)/2$ is the number of parameters and $\hat{\bfbeta}$ and $\hat{\bfSigma}$ is the maximum likelihood estimator of $\bfbeta$ and $\bfSigma$, respectively.
Note that the BIC penalizes more than the AIC the inclusion of parameters in the model, so it tends to choose more parsimonious models.
Smaller values of AIC and BIC indicate better fitted models.

\subsection{Widely available information criterion}

The widely available (Watanabe-Akaike) information criterion (WAIC) was proposed by \cite{watanabe2010}. Compared to AIC and BIC, WAIC averages over the posterior distribution rather than conditioning on a point estimate. 
The criterion is computed as in \cite{Gelman2014}:
\begin{eqnarray*}
	{\rm WAIC} = -2({\rm lppd} - {\rm pWAIC}),
\end{eqnarray*}
where lppd is the log pointwise predictive density, which measures the quality of the model fitting, and is computed as
\begin{eqnarray*}
	\sum_{i=1}^n\log\left( \frac{1}{S}\sum_{s=1}^S l(\bfTheta^s;\bfy_i)\right),
\end{eqnarray*}
with $\Theta^s$, labeling the $s$-th sampled value from the posterior distribution, $s=1,\dots,S$.
The effective number of parameters is computed as
\begin{eqnarray*}
	{\rm pWAIC} = \sum_{i=1}^n V_{s=1}^S \left(\log \, \, l(\bfTheta^s;\bfy_i)\right),
\end{eqnarray*}
with $V_{s=1}^S(\cdot)$ corresponding to the sample variance. As AIC and BIC, smaller values of WAIC indicate better fitted models.

\section{Assessment of MCMC with the MSQ dataset}\label{Ap3}

\begin{figure}[h!]
	\centering
    {\includegraphics[scale=0.53]{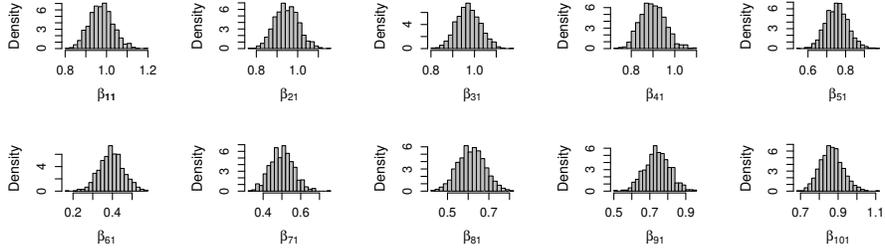}}
	\vspace{-1.4 cm} \caption{Marginal posteriors of components of $\bfbeta$ when fitting a one-factor structure to the MSQ dataset.}\label{fig3}
\end{figure}

\begin{figure}[h!]
	\centering
	{\includegraphics[scale=0.53]{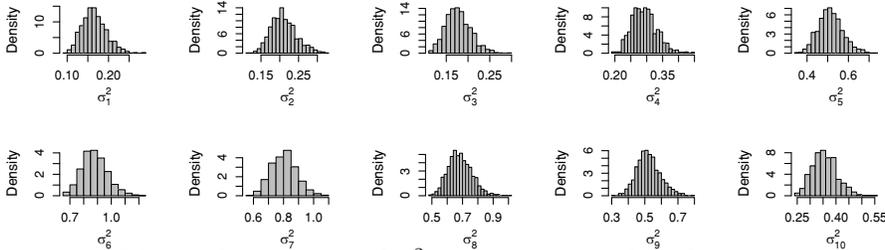}}
\vspace{-0.5 cm} \caption{Marginal posteriors of $\sigma^2_j$ $j=1,\dots,10$  when fitting a one-factor structure to the MSQ dataset.}\label{fig4}
\end{figure}

Figures \ref{fig3} and \ref{fig4} show, respectively, the histograms with the posterior samples of the factor loadings and variances obtained from the analysis of the MSQ dataset using the probit factor model assuming $q=1$. Note that the marginal posteriors graphed are all unimodal. 

\begin{figure}[h!]
	\centering
	{\includegraphics[scale=0.53]{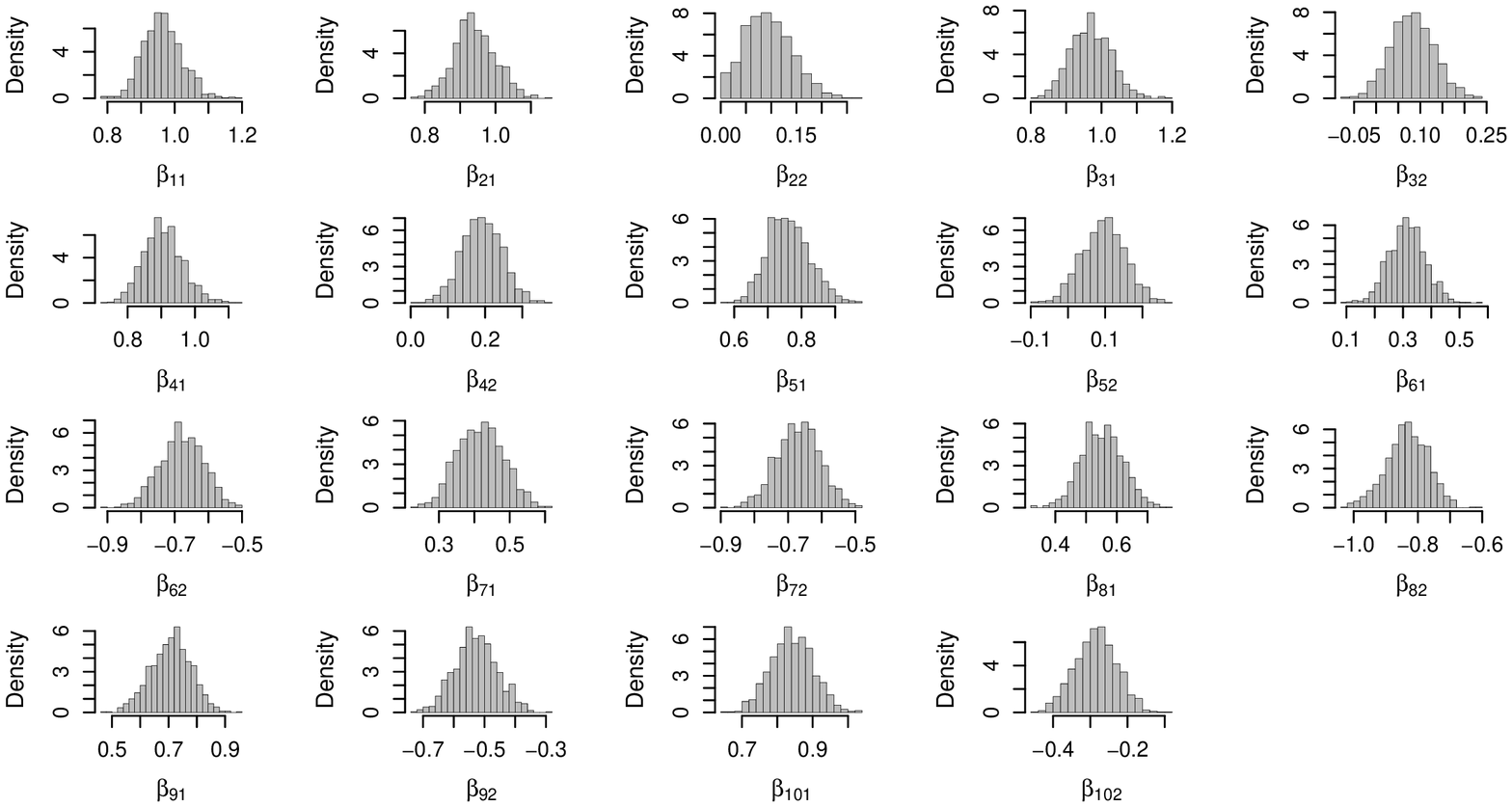}}
	\vspace{-0.8 cm} \caption{Marginal posteriors of components of $\bfbeta$ when fitting a two-factor structure to the MSQ dataset.}\label{fig5}
\end{figure}

\begin{figure}[h!]
	\centering
	{\includegraphics[scale=0.53]{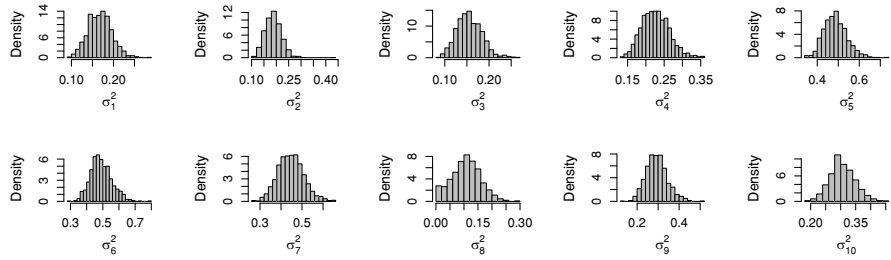}}
	\vspace{-0.8 cm} \caption{Marginal posteriors of $\sigma^2_j$ $j=1,\dots,10$  when fitting a two-factor structure to the MSQ dataset.}\label{fig6}
\end{figure}

Figures \ref{fig5} and \ref{fig6} show, respectively, the histograms with the posterior samples of the factor loadings and variances obtained from the analysis of the MSQ dataset using the probit factor model assuming $q=2$. Note that the marginal posteriors graphed are all unimodal. 

\begin{figure}[h!]
	\centering
	{\includegraphics[scale=0.55 ]{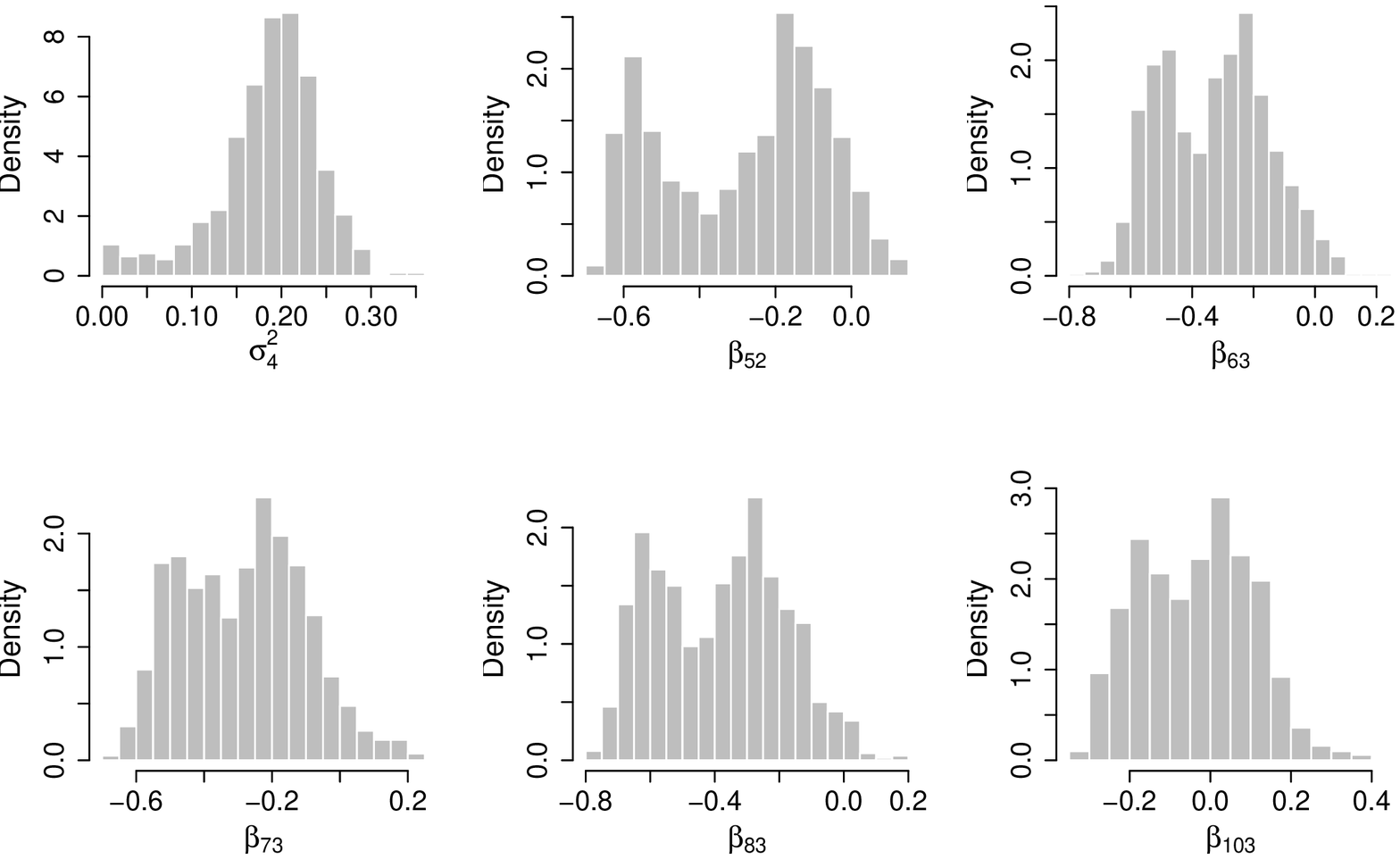}}
	\vspace{-0.8 cm} \caption{Marginal posteriors of some components of $\bfbeta$ and $\bfSigma$ when fitting a three-factor structure to the MSQ dataset.}\label{fig7}
\end{figure}

On the other hand, Figure \ref{fig7} displays histograms with the posterior sample of some of the idiosyncratic variances and components of the factor loadings matrix obtained after fitting the probit model for ordered categories assuming $q=3$ for the MSQ dataset. Note that multiple modes appear in the analysis
of a three-factor model in the same way as seen in Figure \ref{fig2} in Subsection \ref{study2}. Thus, different from the two-factor model, the margins obtained from the three-factor model fit are consistent with the view that $q=3$ is too large in this case, and thus provides additional support for the two-factor model.

\end{document}